\documentclass[a4paper]{JAC2000}
\usepackage{epsf}
\usepackage{daniel}
\title{BEAM DYNAMICS SIMULATION FOR THE CTF3 DRIVE-BEAM ACCELERATOR}
\author{D. Schulte, CERN, 1211 Geneva, Switzerland}
\begin{document}
\maketitle
\begin{abstract}
A new CLIC Test Facility (CTF3) at CERN will serve to study the drive
beam generation for the Compact Linear Collider (CLIC).
CTF3 has to accelerate a $3.5\,\rm A$ electron beam in almost
fully-loaded structures. The pulse contains more than 2000 bunches, one in
every second
RF bucket, and has a length of more than one $\mu s$. Different options
for the lattice of the drive-beam accelerator are presented, based on
FODO-cells and triplets as well as solenoids. The transverse stability is
simulated, including the effects of beam jitter, alignment and beam-based
correction.
\end{abstract}

\section{Introduction}
In the nominal stage of CTF3, the drive-beam accelerator will have eight
klystrons, each feeding two 1\u{m}-long structures.
The structures are almost fully loaded, transferring more than $90\u{\%}$ of
their input power to the beam. The average energy gain per structure is
$\Delta E\approx9.1\u{MeV}$~\cite{c:igor}.
The beam pulse consists of ten short trains of about 210 bunches each.
The first train fills odd buckets, the immediately following second train fills
even buckets; this pattern is then repeated.
An RF-deflector at half
the linac frequency is used to separate the trains after
acceleration~\cite{c:delay}.
The initial beam energy is $E_0\approx26\u{MeV}$, the final beam energy
$E_f\approx170\u{MeV}$, the bunch charge
$q=2.33\u{nC}$, its length $\sigma_x\approx1.5\u{mm}$~\cite{c:louis}
and the transverse normalised emittances are
$\epsilon^*_x=\epsilon^*_y=100\u{\mu m}$.

\section{Structure Model}
The simulations below have been performed using PLACET~\cite{c:placet}.
The long-range transverse wakefield is represented by the
lowest two dipole modes of each cell.
These have been calculated neglecting the coupling between
cells and the effect of the damping waveguides~\cite{c:lars}.
The damping of the lowest dipole mode has been found~\cite{c:michel}
to be in the range $Q=11$ to $Q=19$ for perfect loads.
In the simulation, the modes are confined to their cells, which
allows one to take into account the angle of the beam trajectory
in the structure. The loss factors used in the simulation are
$50\u{\%}$ larger than in~\cite{c:lars}. This is to account for the effect of
higher-order modes. Also, the damping is conservative in the
simulation; $Q=30$ and $Q=400$ are used for the lowest and the
second dipole band. The short-range longitudinal~\cite{c:lars} and
transverse~\cite{c:antonio} wakefields have been calculated and are included
in the simulation. Almost perfect compensation of
the long-range longitudinal wakefields is predicted~\cite{c:igor}.

Quadrupole wakefields may be important and have been implemented in PLACET.
The corresponding modes have not yet been calculated but need to be included
in the simulation as soon as they are available.

\section{Lattices}
Three different lattices were investigated. One consists of simple FODO-cells,
with one structure between each pair of quadrupoles.
The other two lattices are based on triplets. In one case (called T1 below),
one structure was placed between two triplets; in the other case two
structures (T2). The weaker triplet lattice (T1)
and the FODO lattice are roughly comparable in length and cost,
whereas the strong triplet lattice (T2)
is significantly longer and more costly.

In the FODO lattice, the phase advance is $\mu=102^\circ$ per cell, with a
quadrupole spacing of $2\u{m}$. In T2 one has
$\mu_x=97^\circ$ and $\mu_y=110^\circ$, and a distance of $4.2\u{m}$ between
triplets. The sum of the integrated strengths of the outer two
magnets is slightly larger than that of the inner one. With
this arrangement, the horizontal and the vertical beta-functions are equal in
the accelerating structures, and the energy acceptance of the lattice is
markedly improved. For T1 the phase advances are $\mu_x=84^\circ$ and
$\mu_y=108^\circ$ for a triplet spacing of $3\u{m}$.
The transverse
acceptance is $4.2\,\sigma$ for the FODO lattice, $4.9\,\sigma$ for T2 and
$5.8\,\sigma$ for T1.

Since the beams have to be compressed after the acceleration, the RF-phase
cannot be used to optimise the beam transport. It must be chosen to
achieve the required compression and to limit the energy spread of the
beam before the combiner ring to the latter's energy acceptance.
An RF phase $\Phi_{RF}=6^\circ$ is used in the following.

\section{Transverse Beam Jitter}
\begin{figure}
\epsfxsize=4cm
\epsfbox{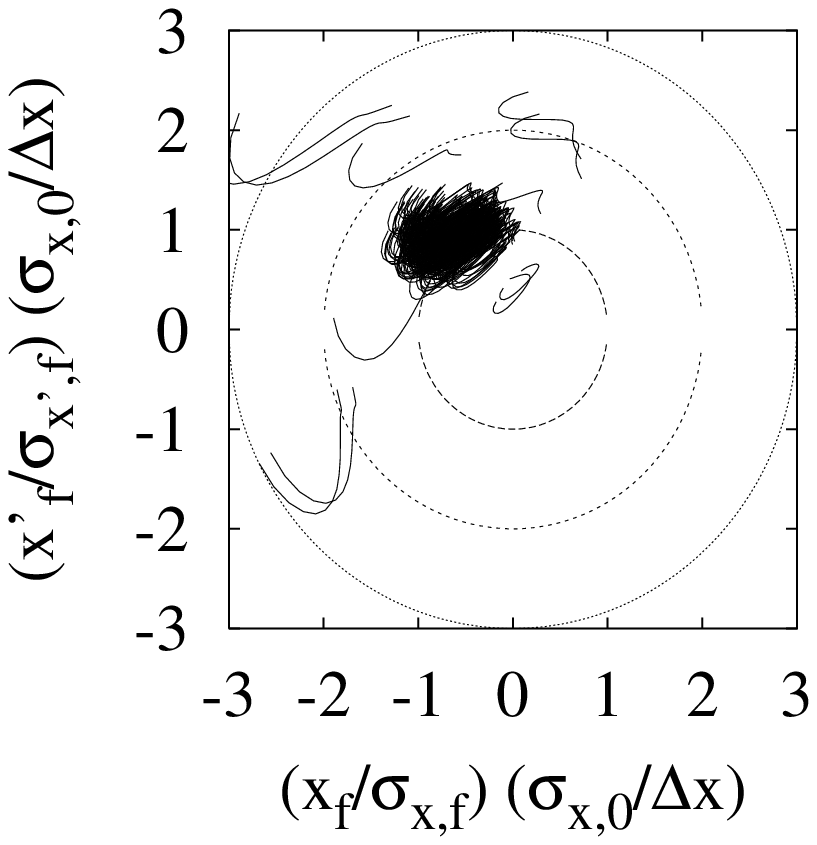}
\epsfxsize=4cm
\epsfbox{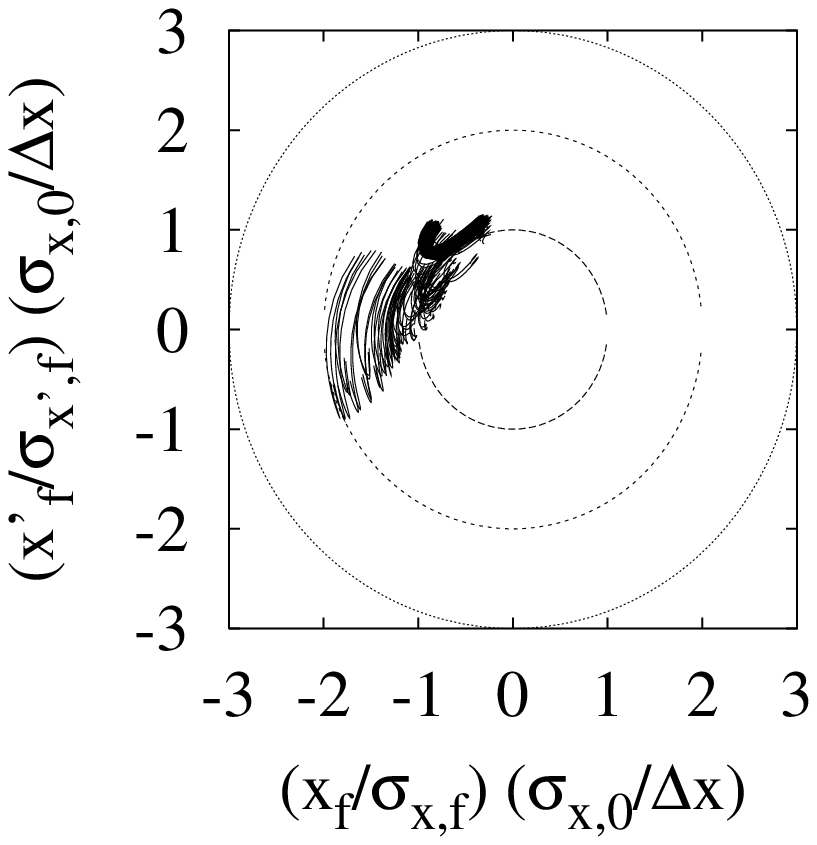}
\caption{The amplifiaction factor of the beam at the end of the drive-beam
accelerator, using the FODO lattice,
without a ramp (left) and with a ramp (right).
A mono-energetic beam without wakefields should stay on the
innermost circle.}
\label{f:jitt1}
\end{figure}

No estimate of the transverse jitter of the incoming beam exists.
Therefore, only the jitter amplification is calculated. In the simulation,
each bunch is cut into slices; the beam is set to an offset of
$\Delta x$ and tracked through the linac. The normalised amplification
factor $A$ for a slice is defined as
\begin{displaymath}
A=\frac{\sigma_{x,0}}{\Delta x}
\sqrt{\left(\frac{x_f}{\sigma_{x,f}}\right)^2
+\left(\frac{x^\prime_f}{\sigma_{x^\prime,f}}\right)^2}
\end{displaymath}
Here, $\sigma_{x,0}$ and $\sigma_{x,f}$ are initial and final beam size,
$\sigma_{x^\prime,0}$ and $\sigma_{x^\prime,f}$ are initial and final beam
divergence, $\Delta x$ is the initial beam offset and $x_f$ and $x^\prime_f$
are the final position and angle of the centre of the slice.
For a slice with nominal energy and without wakefield effects, one has $A=1$.
The maximum amplification factor $\hat A$ is the maximum over all slices.
The left-hand side of Fig.~\ref{f:jitt1} shows the bunches at the end of the
accelerator using the FODO lattice.
Different quadrupole strengths were used to find the best phase advance.
Some bunches are kicked significantly; the maximum amplification is
$\hat A=3.7$.
Without knowledge of the acceptance downstream and the size of the beam jitter,
it is not possible to decide whether the amplification is acceptable.
Within the linac, even a large jitter of
$\Delta x=\sigma_x$ does not lead to beam loss.

\begin{figure}
\epsfxsize=4cm
\epsfbox{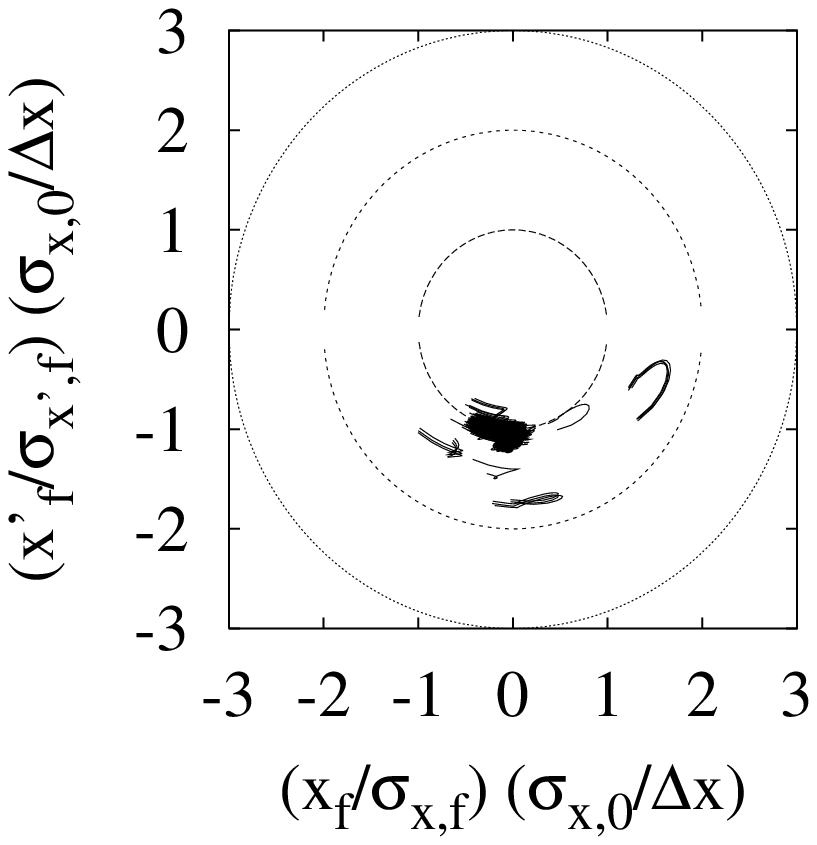}
\epsfxsize=4cm
\epsfbox{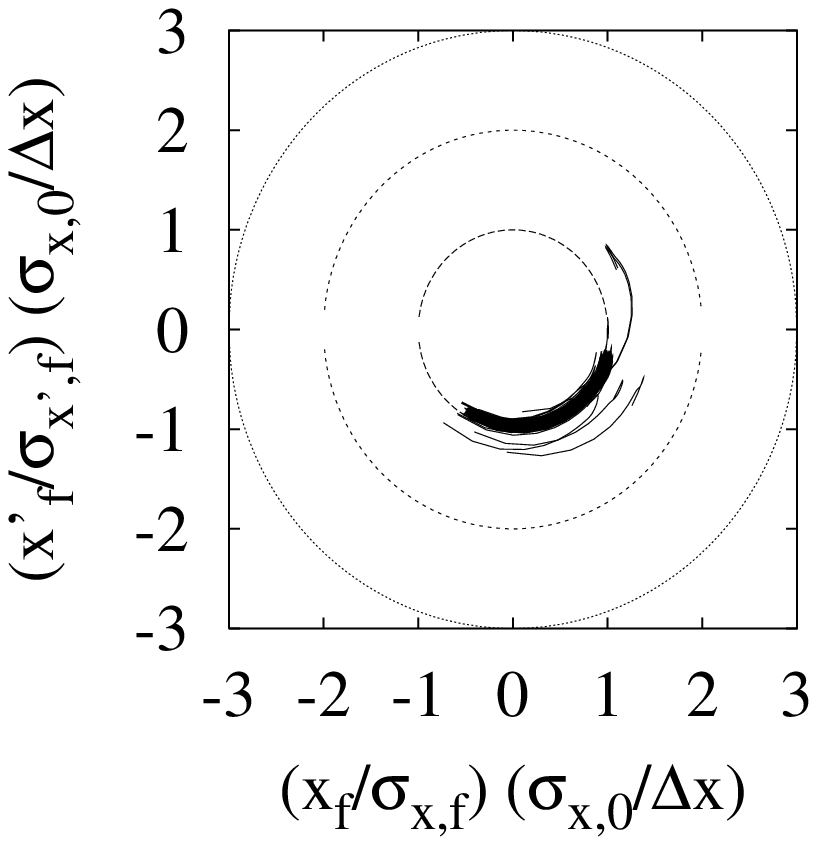}
\caption{The beam at the end of the drive-beam accelerator in a triplet
lattice. On the left-hand side T2,
on the right-hand side T1.}
\label{f:jitt3}
\end{figure}

The first few bunches in each train are kicked particularly hard. This can be
prevented by adding charge ramps.
Towards the end of a train that fills even buckets, the bunch charge is slowly
decreased from the nominal bunch charge to zero.
At the same time one increases the charge in the odd buckets from zero to
nominal, to keep the beam current constant. Thus the two consecutive trains
practically overlap.
On the right-hand side of Fig.~\ref{f:jitt1}, one can see that in this case
all bunches are well confined, with a maximum amplification of $\hat A=2$.

In the triplet lattices, the horizontal plane has a larger jitter amplification
than the vertical one. But even the horizontal amplifications are
significantly smaller than in the FODO lattice.
Figure~\ref{f:jitt3} shows the examples of a pulse
without charge ramps, the amplification factors being 1.8 (T2) and 1.5 (T1).
With charge ramps, they are reduced to 1.5 and 1.3.
If the beam jitters significantly, the triplet lattices are markedly
better than the FODO lattice. 

\section{Beam-Based Alignment}
To keep operation as simple as possible, only
one-to-one correction is considered. All elements are assumed to be scattered
around a straight line following a normal distribution with
$\sigma=200\u{\mu m}$.
In the FODO lattice, corrector dipoles are located after each
quadrupole and beam position monitors (BPM) are placed in front of each
quadrupole.
In the triplet lattices, the corrector dipoles are positioned after the
triplets and the BPMs are positioned in front and after the triplets.
The correctors are used to bring the average beam position to zero in the BPMs.
For each case, 100 different machines are
simulated. The small growths of about $0.5\u{\%}$ are almost the same for
all lattices.

\section{Gradient and Phase Errors}
The limit on the variation of the bunch energy is
$1\u{\%}$~\cite{c:roberto}, much smaller than the single-bunch energy spread.
In normal operation, the additional dispersive effects are therefore small.
Static local energy errors are of more concern and are discussed here.

\begin{figure}
\epsfxsize=8cm
\epsfbox{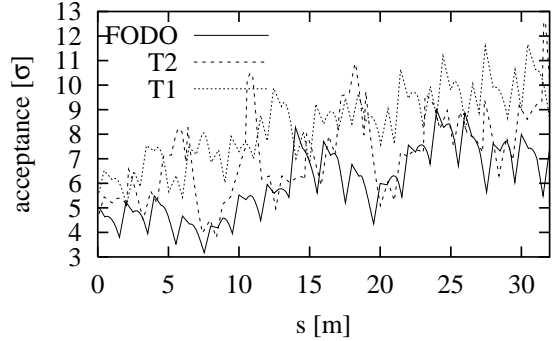}
\caption{The minimum acceptance along the linac, with a gradient error.
For each lattice, 100 machines have been simulated and their minimum
acceptance at each point is plotted.}
\label{f:energy}
\end{figure}
The initial and final beam energy can be well measured, and from this
the average gradient can be derived. A local variation of the gradient is
more difficult to detect. It will lead to a quadrupole strength that is
not adapted to the beam energy. The worst case is too low a gradient in the
first two structures, which are fed by one klystron.
In the simulation, 100 different machines with
a gradient in the first two structures that is too low by $10\u{\%}$
($20\u{\%}$) are corrected with the beam.
The emittance growth found after correction is $1\u{\%}$ ($5\u{\%}$)
in the FODO lattice
and $0.5\u{\%}$ ($2\u{\%}$) in T1, which seems to be sufficiently low.
In T2, the value for a $10\u{\%}$ error is small, $2\u{\%}$,
but for an error of $20\u{\%}$ it starts to be large: $14\u{\%}$.
The transverse acceptance is reduced to
$3.8\,\sigma$ ($3.2\,\sigma$) in the FODO lattice, $4.7\,\sigma$
($3.8\,\sigma$) in T2 and to
$5.2\,\sigma$ ($5.1\,\sigma$) in T1.
Figure~\ref{f:energy} shows the acceptance for a gradient error of
$20\u{\%}$.
For the FODO lattice, and to a lesser degree also for T2, one starts
to worry about beam losses.
However, an error of $10\u{\%}$ seems acceptable with all lattices.
To be able to use the FODO lattice or T2,
it necessary to measure the local gradient to better than $10\u{\%}$,
to be able to correct the lattice accordingly. For T1, a precision of
$20\u{\%}$ is sufficient.

The RF power produced by a klystron has a systematic phase variation during
the pulse.
One hopes to correct this effect globally, but local variations will remain.
To estimate their importance, a linear change in phase of $20^\circ$
over the pulse is assumed for the two structures driven by one klystron.
The next pair has an exactly opposite phase variation.
The resulting bunch-to-bunch energy spread is $2\u{\%}$, full width, which is
not acceptable in the combiner ring; so a better compensation would be needed.
In contrast, the emittance growth seems acceptable with
about $1.5\u{\%}$ averaged over 100 machines for all lattices; the
acceptance is hardly decreased. This phase
variation does not cause significant transverse effects.

\begin{figure}
\epsfxsize=8cm
\epsfbox{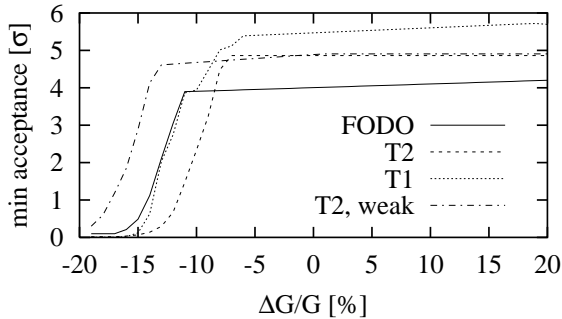}
\caption{The minimum acceptance of the linac, as a function of the RF-gradient
error; 20 machines were simulated for each case.}
\label{f:phase}
\end{figure}

\begin{figure}
\epsfxsize=4cm
\epsfbox{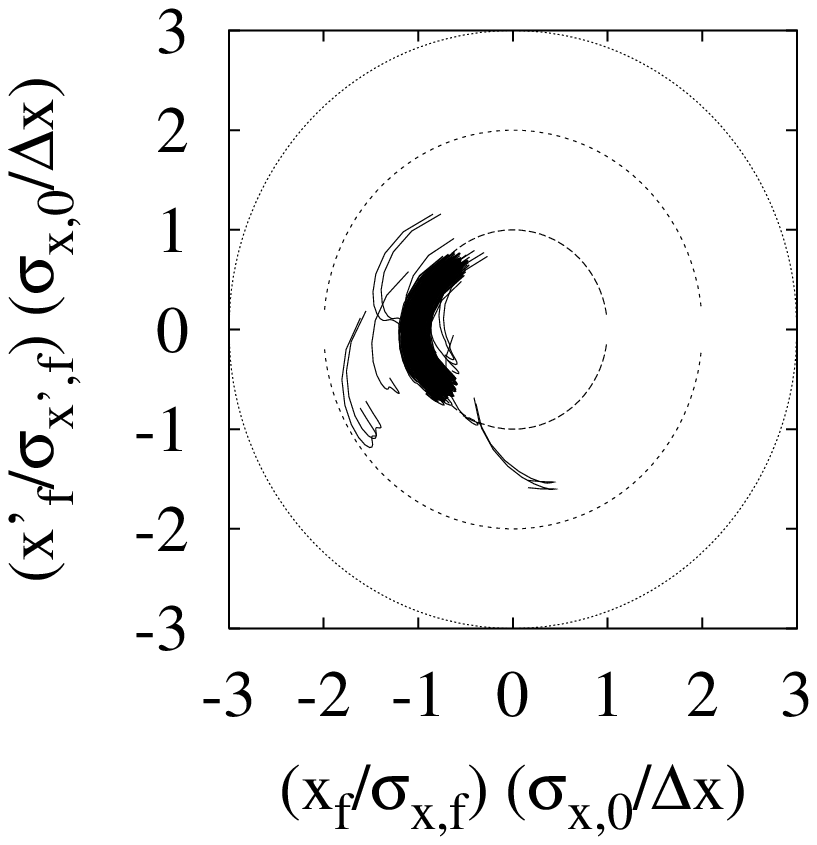}
\epsfxsize=4cm
\epsfbox{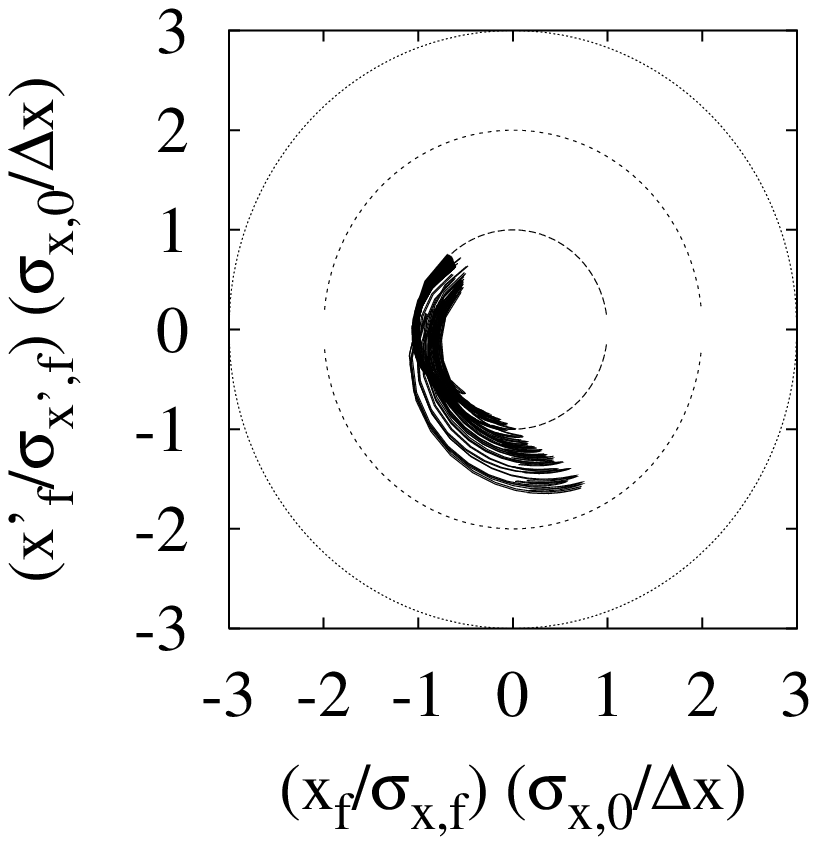}
\caption{The amplification of beam jitter with and without the
charge ramp for the lattice T2 including the injector solenoid.}
\label{f:solenoid}
\end{figure}

\section{Energy Acceptance}
During commissioning of the linac, large energy errors may occur.
To study the sensitivity to this, 20 machines were simulated for each
lattice in the following way: the linac is corrected
with a nominal beam; then the incoming beam is assumed to be accelerated at a
different gradient.
Figure~\ref{f:phase} shows the minimum transverse acceptance of the three
lattices as a function of the RF-gradient error. The final
energy error is about $1.4$ times larger than that of the RF-gradient,
since the beam loading does not change.
The FODO lattice and T1 have a comparable energy acceptance, whereas
that of T2 is slightly smaller.
By reducing the focal strength, the energy acceptance can be further
improved at the cost of higher transverse wakefield effects.
By reducing the strength of T2 to $\mu_x=83^\circ$ and $\mu_y=94^\circ$,
the energy acceptance becomes larger than that of the FODO lattice. The maximum
amplification of an initial jitter increases from 1.8 to 2.6 but is still
smaller than the factor 3.7 in the FODO lattice.
With each lattice, the linac energy acceptance is largely sufficient
during normal operation.

\section{Solenoid}
PLACET has been modified to also simulate the effects of
solenoids with acceleration.
This allows to include the last two structures of the injector which are
placed inside a solenoid. Two triplets are used to match the end of the
solenoid to the T2 version of
the drive-beam accelerator. The field of the solenoid is $0.2\u{T}$ and
its length is chosen such that a horizontal jitter
of the nominal beam leads to a final horizontal offset.
The end fields of the solenoid are modelled as thin lenses.
Neither space charge nor the difference of the particle velocities from
the speed of light are taken into account, but the wakefields are considered,
in contrast to calculations done with PARMELA~\cite{c:feng}.

Figure~\ref{f:solenoid} shows the amplification factor. While there is some
contribution from the structures in the solenoid, the overall amplification
seems still acceptable.

\section{Conclusion}
The simulations show that the lattices considered here can be acceptable;
the best is the strong triplet lattice T1. The triplet lattice T2 seems
to be a better choice than the FODO lattice. The FODO lattice is
less expensive than T2, which is much cheaper than T1.
To find the best compromise, more information is needed.
For the FODO lattice the ramps have to be studied in more detail.
For all lattices, the matching from the injector to the linac and from the
linac to the combiner ring needs to be understood.


\begin{thebibliography}{10}

\bibitem{c:igor} I. Syratchev. Private communication.

\bibitem{c:delay} D. Schulte.
The Drive-Beam Accelerator of CLIC. {\it Proceedings of Linac 1998,
Chicago, USA} and {\it CERN/PS 98-042 (LP)} (1998).

\bibitem{c:louis} L. Rinolfi. Private communication.

\bibitem{c:placet} D. Schulte. PLACET: A Program to Simulate Drive Beams.
{\it Proceeding of EPAC 2000, Wien, Austria} and
{\it CERN-PS-2000-028 (AE)} (2000).

\bibitem{c:lars} L. Thorndahl. In: The CLIC RF Power Source.
{\it CERN 99-06} (1999).

\bibitem{c:michel} E.~Jensen, A.~Millich and L.~Thorndahl.
Private communication.

\bibitem{c:antonio} A. Millich. Private communication.

\bibitem{c:roberto} R. Corsini. Private communication.

\bibitem{c:feng} F. Zhou. To be published as a CTF3-Note.
\end{thebibliography}
\end{document}